\documentclass[aps,preprint,showpacs]{revtex4-1}
\usepackage{graphicx,subfigure}
\usepackage{amsmath,amsfonts,amssymb}
\usepackage{hyperref}

\newcommand{\f}{\frac}
\newcommand{\suml}{\sum\limits}

\graphicspath{{D:/localtexmf/Mytex/LGH/mathplots/}}
\begin{document}
\title[Surface tension]{The application of the global isomorphism to the surface tension of the liquid-vapor interface of the Lennard-Jones fluids}
\author{V.~Kulinskii}
\email{kulinskij@onu.edu.ua}
\affiliation{Department for Theoretical
Physics, Odessa National University, Dvoryanskaya 2, 65026 Odessa, Ukraine}
%%%%%%%%%%%%%%%%%%%%%%%%%%%%%%%%%%%%%%%%%%%%%%%%%%%%%%%%%%%%%%%%%%%%%
\begin{abstract}
In this communication we show that the surface tension of the real fluids of the Lennard-Jones type can be obtained from the surface tension of the lattice gas (Ising model) on the basis of the global isomorphism approach developed earlier for the bulk properties.
\end{abstract}
 \maketitle
%\section{Introduction}
The lattice models play the important role in the understanding the behavior of the real systems which are too complex to be treated in controllable way. In general the direct comparison between real continuum systems and their discrete models is impossible. That is why the situations where one may to
establish the quantitative relations between the physical properties of these systems are of great importance. Recently,
the conception of the global isomorphism between Lennard-Jones fluids and the short-ranged Ising model (lattice gas) was proposed in \cite{eos_zenomeglobal_jcp2010}. It is based on the simple geometrical reformulation in \cite{eos_zenome_jphyschemb2010} the results of the profound analysis of the thermodynamic data for real and model systems made by E.~Apfelbaum and V.~Vorob'ev in a series of papers \cite{eos_zenoapfelbaum_jpchemb2006,eos_zenoapfelbaum1_jpcb2009}. These authors stressed the necessity to pay special attention  on the linear regularities characteristic for the molecular fluids known more than a century. These linearities are the rectilinear diameter law \cite{crit_diam0} and, especially, the Zeno line linearity \cite{eos_zenoholleran_jpc1969,eos_zenoboyle_jphysc1983,eos_zenobenamotz_isrchemphysj1990} in the simplified form known as the Batchinsky law \cite{eos_zenobatschinski_annphys1906}. In \cite{eos_zenome_jphyschemb2010} it was shown how these phenomenology could be casted in clear geometrical picture via the mapping between the set of the states of the fluid and the lattice gas of the form:

The objective of this paper is to use the global isomorphism
transformation to relate the surface tension of the liquid-vapor interface of the lattice gas (Ising model) with that of the LJ fluid. The results obtained for the binodal in \cite{crit_globalisome_jcp2010} give the ground to use the approach for the surface tension since from the thermodynamical point of view this physical quantity is nothing but the thermodynamic potential of the surface. It has obvious meaning in the lattice gas model and due to the existence of the exact solutions it is possible to introduce the thickness of the surface.

\begin{figure}
\center
  \includegraphics[scale=1]{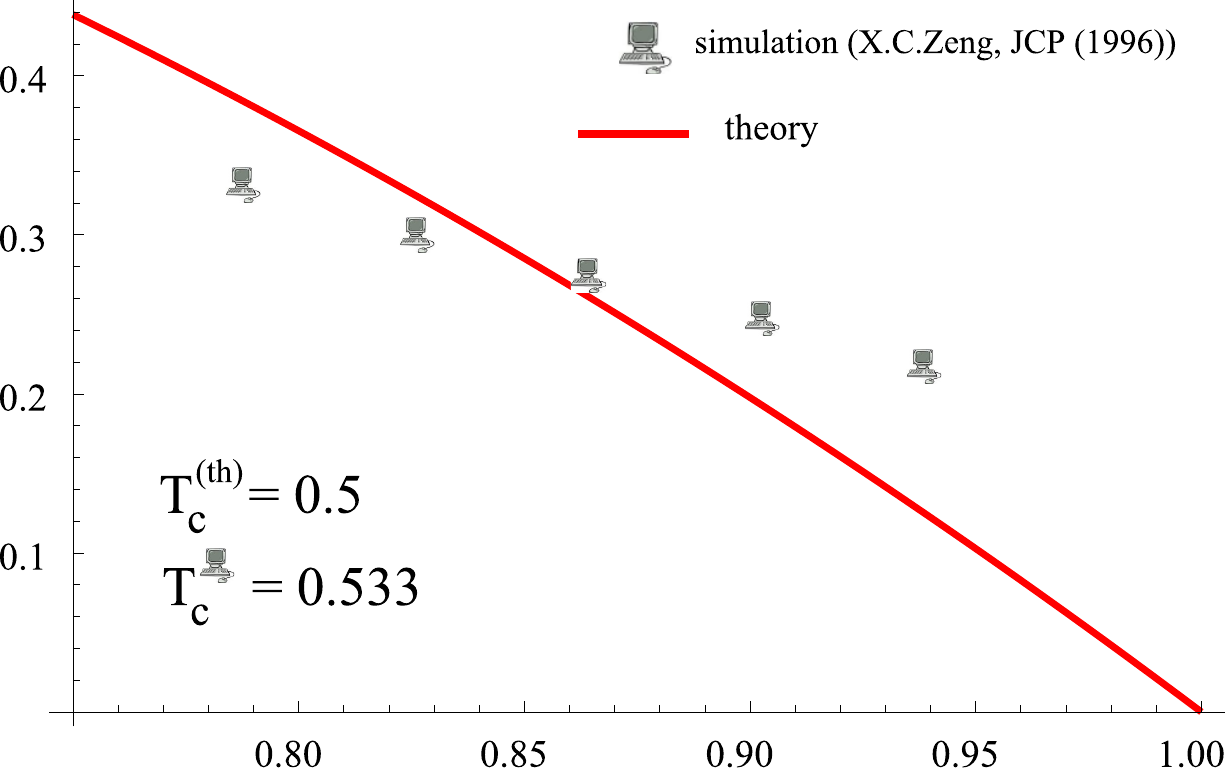}\\
  \caption{The comparison of the surface tension of the 2D  Lennard-Jones fluid given by Eq.~\eqref{st_2dlj}
  with the simulation results of
  \cite{crit_2dsurftens_jcp1996}.}\label{fig_surftens_2dcomparis}
\end{figure}

\begin{figure}
\center
\includegraphics[scale=0.75]{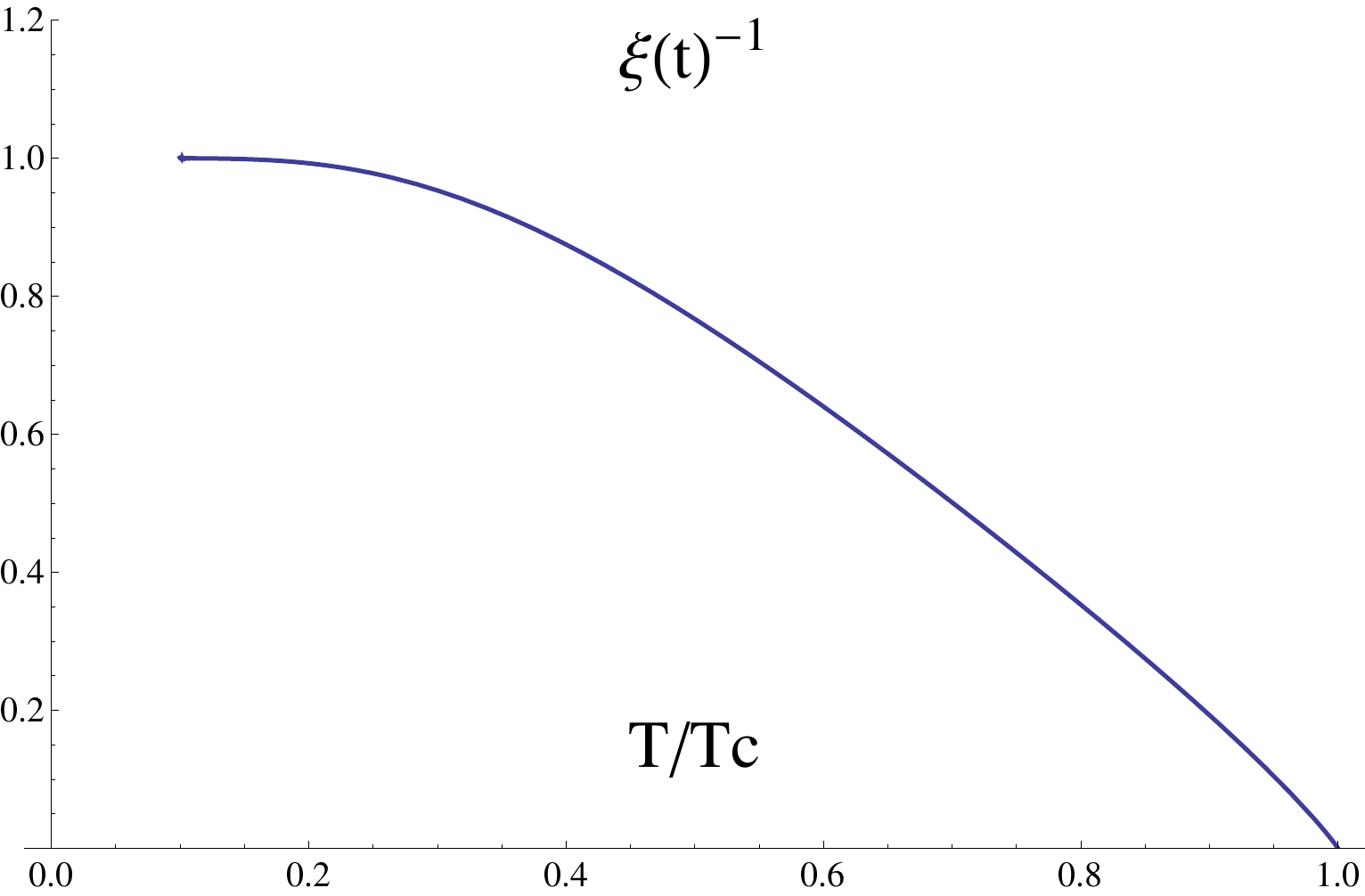}
\caption{Effective interfacial thickness for $2D$ Ising model from Eq.~\eqref{st_woodmodif} with $\sigma_0 = 1/4, \eta = 1/4$ on the basis of the Onsager's solution Eq.~\eqref{ising2donsager}.}\label{fig_ising2dthickness}
\end{figure}

The following mapping representing the global isomorphism between the thermodynamic states of the LG and the LJ fluid was constructed:
\begin{equation}\label{projtransfr}
  n =\, n_{*}\,\f{x}{1+a\,t}\,,\quad
  T =\, T_{*}\,\f{a\, \tilde{t}}{1+a\,\tilde{t}}\,,
\end{equation}
with
\[a = \f{T_c}{T_{*}-T_c}\,.\]
Here $n$ and $T$ are the density and the temperature of the fluid, $\tilde{t}$ is the temperature variable of the LG normalized to the critical temperature so that $\tilde{t}_c = 1$. We also use the standard dimensionless values for $T$ and $n$ of the LJ fluid \cite{book_hansenmcdonald}. $T_{*}$ and $n_{*}$ are the parameters of the linear zeno-element:
\begin{equation}\label{vdw_z1}
  \f{n}{n_{*}}+\f{T}{T_{*}} = 1\,.
\end{equation}
The coordinates of the CP for the fluid are:
\begin{equation}\label{cp_fluid}
  n_{c} = \f{n_{*}}{2\left(\,1+a\,\right)}\,,\quad   T_{c} = T_{*}\, \f{a}{1+a}\,.
\end{equation}

\[\Sigma^{(lat)}_{\mathcal{N}} = \Lambda^{m_{2}}_{max}+ \Lambda^{m_{2}}_{1} + \ldots\]
where $\mathcal{N} = m_{1}\times m_{2}$ is the size of the lattice. The thermodynamic potential is augmented with the surface contribution is:
\begin{equation}\label{globaliso_potentials}
V\,T\,\ln\,\Xi_{V}(\mu,T) = V\,P+\sigma\,A = \mathcal{N}\,\mathfrak{g} + \mathfrak{s}\,\mathcal{A} = \mathcal{N}\,t\,\ln \Sigma_{\mathcal{N}}(h,t)
\end{equation}
\begin{equation}\label{st_2dising}
\mathfrak{s}(t) = 2 + t\,\ln
\left(\,\tanh\,\frac{1}{t} \,\right) = 4\,|\tau|+o(\tau)\,,\quad \tau = 1-t/t_c\,.
\end{equation}
Therefore, according to the relation Eq.~\eqref{projtransfr} and Eq.~\eqref{globaliso_potentials} with $z=1/3$ for two-dimensional case this leads to the following result for the surface tension of the liquid-vapor interface of the 2D LJ fluid:
\begin{equation}\label{st_2dlj}
  \sigma_{LJ}(T) = \mathfrak{s}(t(T))  \underset{T\to T_c-0}{=}  \f{16}{3}\,(1-T/T_c)+\ldots\,,
\end{equation}
\begin{figure}
\center
  \includegraphics[scale=0.75]{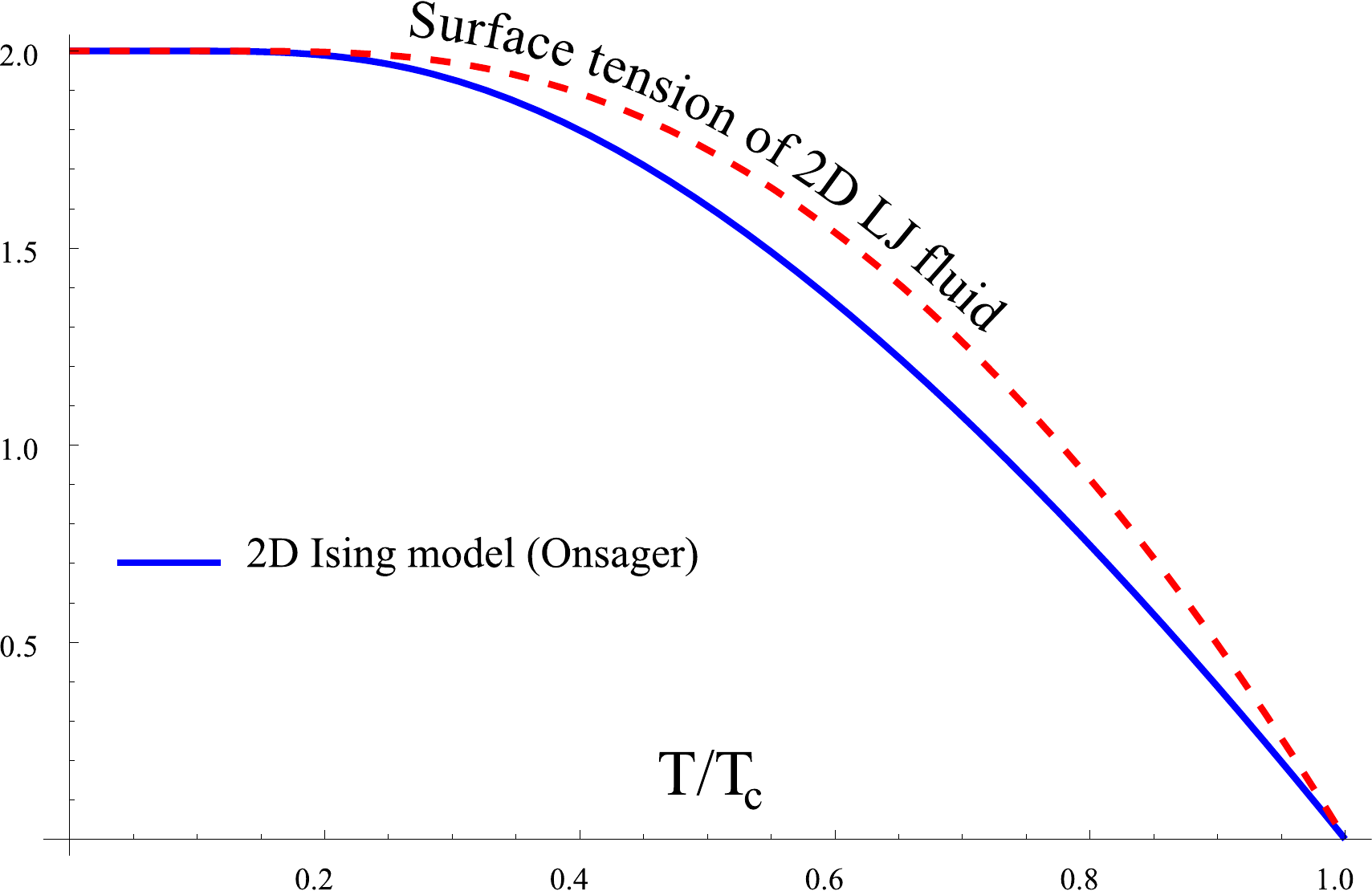}\\
  \caption{The surface tension of the 2D Ising model (Onsager solution) and the surface tension of the 2D Lennard-Jones fluid obtained according to Eq.~\eqref{st_2dlj}.}\label{fig_st_2d}
\end{figure}
The result of the surface tension for the 2D LJ fluid is on Fig.~\ref{fig_st_2d}.
The comparison with the result of the available simulational data for LJ fluid is shown on Fig.~\ref{fig_surftens_2dcomparis}.
In the cases where no exact solution is available one may use the results of \cite{crit_surftensionlatticegas_jcp1970,crit_surftensionlatticegas_jcp1972}. There it was shown that the surface tension of the lattice model can be represented as the difference between averages of the local physical quantity in coexisting phases:
\begin{equation}\label{st_wood}
  \sigma = t\,
\left(\, \left\langle\, \mathfrak{s} \,\right\rangle_{gas} - \left\langle\, \mathfrak{s} \,\right\rangle_{liq} \,\right)\,.
\end{equation}
The representation \eqref{st_wood} is based on the markovian property of the distribution function of the lattice model with the nearest-neighbor interaction. The distribution function of the $D$-dimensional lattice is represented as the markovian chain for $D-1$-dimensional slices. Such representation is widely used in the transfer matrix method \cite{book_baxterexact}.
Here $\mathfrak{s}$ is the local variable which corresponds to the spin distribution in the slice of the gas phase far from the interface. In the simplest mean-field approximation of the Bragg-Williams $\mathfrak{s}$ is reduced to:
\begin{equation}\label{eta_bw}
  \mathfrak{s} = \f{1}{2}\,\suml_{i}\,\ln{p(s_i)}
\end{equation}
where $p(s_i)$ is the distribution of the $i$-th spin in the slice. Substitution Eq.~\eqref{eta_bw} into Eq.~\eqref{st_wood} and taking into account the symmetry between gas the liquid density: $x_{gas}=1-x_{liq}$ we get:
\begin{equation}\label{st_bwood}
  \sigma = \f{t}{2\,l_0^{D-1}}\,\left(\, x_{liq} - x_{gas}\,\right)\,\ln\,\f{x_{liq}}{x_{gas}} =
  \f{t}{2\,l_0^{D-1}}\,\left(\, 1-2\,x_{gas}\,\right)\,\ln\,\f{1-x_{gas}}{x_{gas}}
\end{equation}
or in terms of the magnetization $m  = 2x-1$:
\begin{equation}\label{st_bwood1}
  \sigma = \f{t}{2\,l_0^{D-1}}\,|m(t)|\,\ln\,\f{1+m(t)}{1-m(t)}
\end{equation}
where $l_0$ is the lattice spacing. Further we put $l_0=1$ for simplicity. Sure neglecting the correlation leads to the very crude approximation especially in the fluctuational region so the direct comparison of Eq.~\eqref{st_wood} with the surface data leads to the big difference \cite{crit_surftenslatticegas_jcp1969}.
In order to correct such weakness but conserving the analytical simplicity for the connection between $\mathfrak{s}$ and $x$ we modify Eq.~\eqref{st_bwood1} as following:
\begin{equation}\label{st_woodmodif}
  \sigma = \sigma_0\,\f{t}{\xi^{1-\eta}_{eff}}\,|m(t)|\,\ln\,\f{1+m(t)}{1-m(t)}\,,
\end{equation}
where $\xi_{eff}$ is the effective width of the interface defined by Eq.~\eqref{st_woodmodif}. It should be of order of correlation length. The value $\sigma_0$ is determined by the correspondence with some reference point. For example in case of the 2D Ising model with:
\begin{equation}\label{ising2donsager}
  m(t) =  \left(1-\f{1}{\sinh^4\left(\,2/t \,\right)}\right)^{1/8}
\end{equation}
the value of $\sigma_0$ is determined by the natural asymptote $\xi/a \to 1$ at $t \to 0$  since the thermal fluctuations vanish and the interfacial thickness goes to its minimal value. This allows to consider Eq.~\eqref{st_woodmodif} as the definition of the interfacial width. The corresponding result is on Fig.~\ref{fig_ising2dthickness}. Of course such definition relies on the approximation for $\mathfrak{s}$. But we note that near the critical point Eq.~\eqref{st_woodmodif} leads to:
\begin{equation}\label{st_woodmodif_cp}
  \sigma \propto \f{\left(\, x_{liq} - x_{gas}\,\right)^2
}{\xi^{1-\eta}_{eff}}\,.
\end{equation}
This has the correct critical asymptote for the surface tension $\sigma \propto |\tau|^{(D-1)\,\nu}$. With account of the global isomorphism relations Eq.~\eqref{projtransfr} this transforms to
the dependence $\sigma \propto (n_{liq}-n_{gas})^2$ derived in \cite{liq_ljtailcorr_molphys1995}. So the approximation Eq.~\eqref{eta_bw} is sufficiently good to take into account fluctuations at least in the leading order.

\begin{figure}
\center
\includegraphics[scale=0.5]{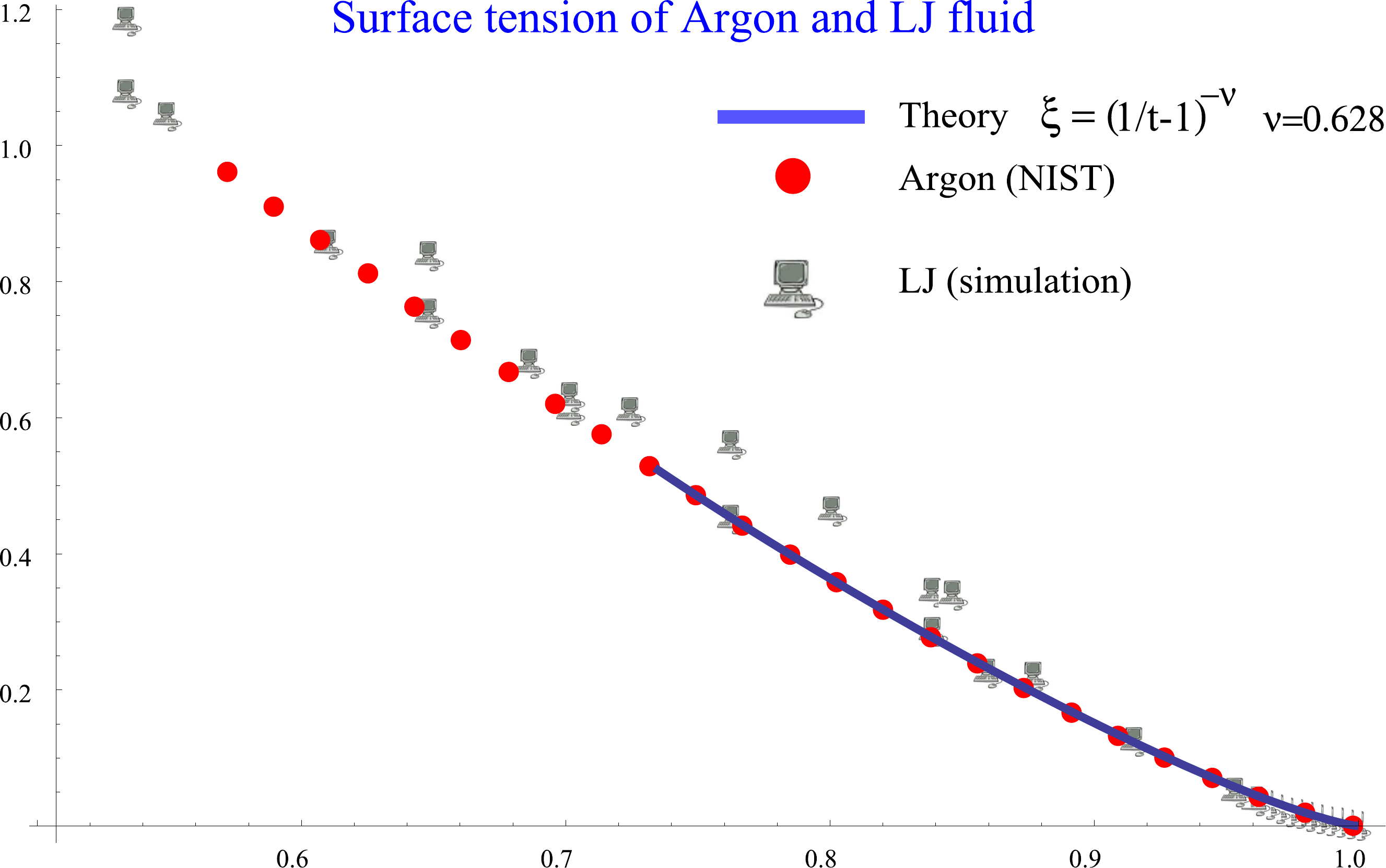}
\caption{Surface tension of the 3D LJ fluid, the argon and the fitting according to Eq.~\eqref{st_real} and Eq.~\eqref{guggenheim_binodal}.}\label{fig_stargon}
\end{figure}

\begin{figure}
\center
\subfigure[]{\includegraphics[scale=0.4]{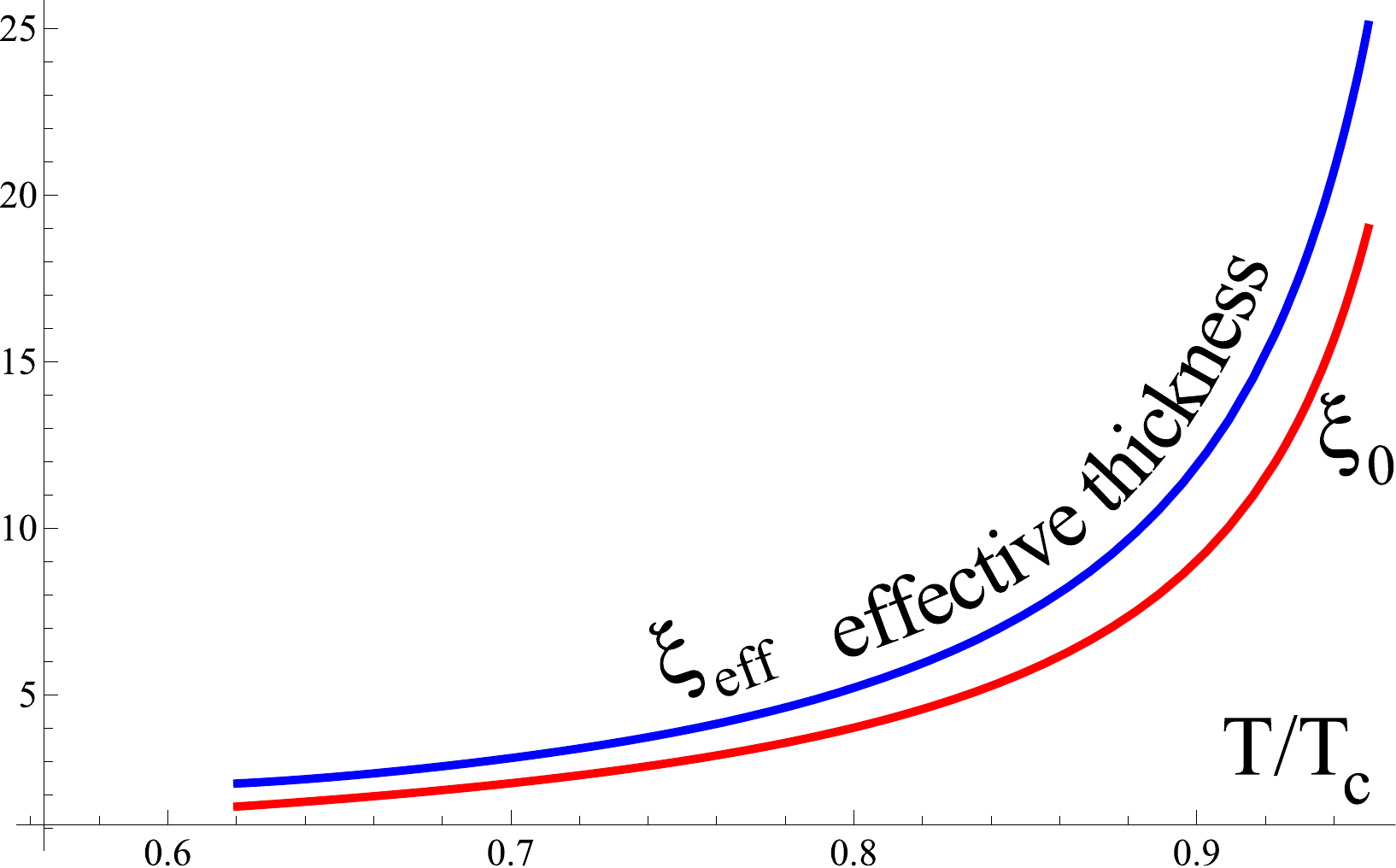}}
\,\,
\subfigure[\,\,$\xi_{eff}/\xi_0$]{\label{fig_xieffcompar3d2}\includegraphics[scale=0.4]{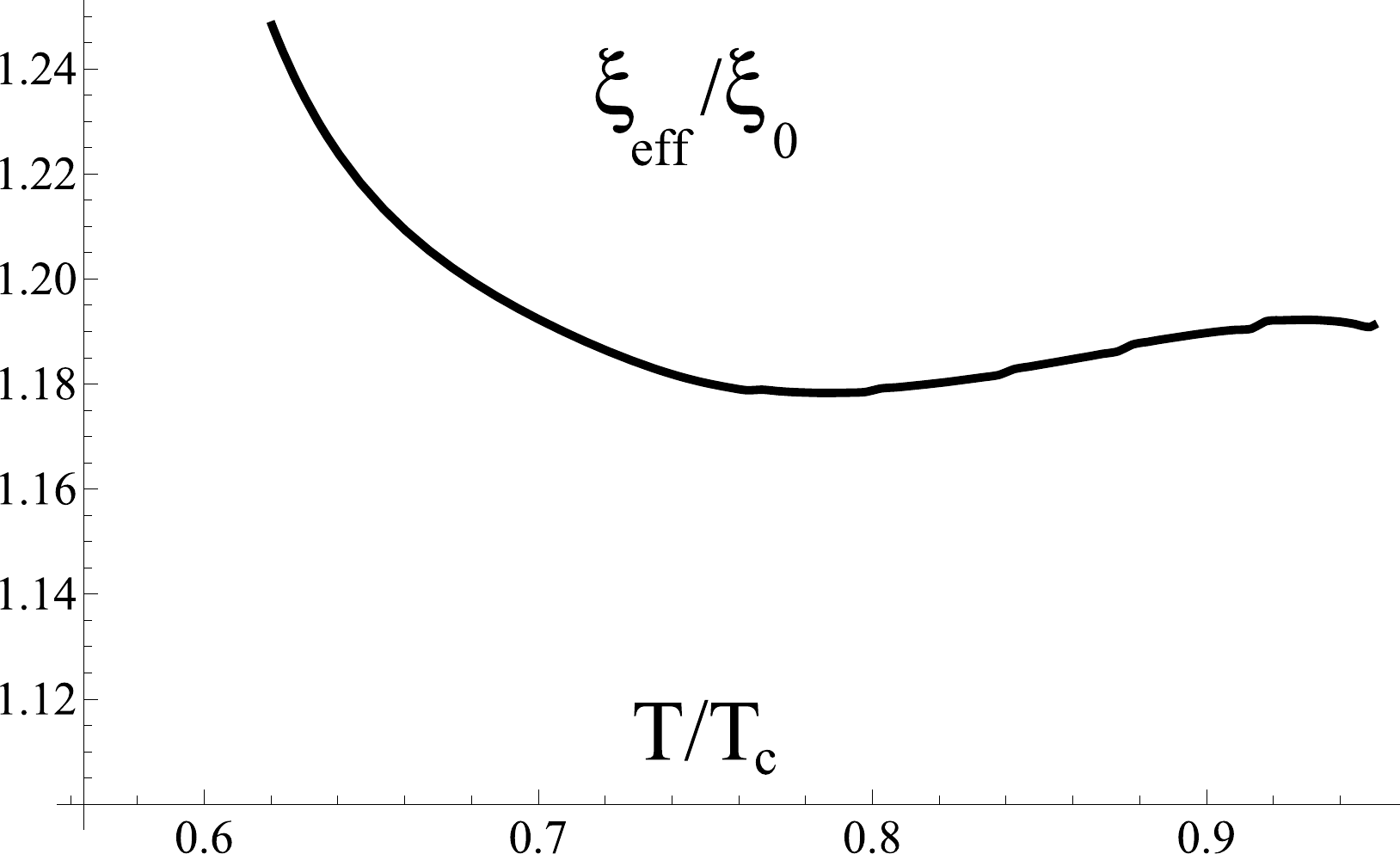}}
\caption{Temperature dependence of the effective interfacial thickness $\xi_{eff}$ for 3D LJ fluid (a) and the ratio $\xi_{eff}$ to simple expression $\xi_0(T)$ (b).}\label{fig_xieffcompar3d}
\end{figure}

Basing on the previous results on the binodals \cite{crit_globalisome_jcp2010} and knowing that in $3D$ case
the fluctuations are less in comparison with $2D$ case we can use the model Eq.~\eqref{st_woodmodif} to reproduce the data for the surface tension of the LJ fluid from the information of the binodal. Using the basic relations Eq.~\eqref{projtransfr} we get:
\begin{equation}\label{st_real}
  \sigma = \sigma_0\,\f{t(T)}{\xi_{eff}^{1-\nu}}\,\f{n_{liq}-n_{gas}}{n_c}\,
  \ln{\f{n_{liq}}{n_{gas}}}\,.
\end{equation}
Here $n_{liq,gas}$ are the densities of the coexisting phases along the binodal for which we use the Guggenheim binodal expression:
\begin{equation}\label{guggenheim_binodal}
  \f{n_{liq,gas}}{n_c} = 1\pm \f{3}{4}\,
\left(\,1-T/T_c\,\right)+\f{7}{4}\,\left(\,1-T/T_c\,\right)^{1/3}\,.
\end{equation}
We also use simple expression for the temperature dependence of the correlation length $\xi_{eff} \to \xi_0 = (T_c/T-1)^{-\nu}$, $\eta\approx 0.03$ - Fisher's critical exponent and the exponent $\nu$ is taken as the fitting parameter. The result are shown on Fig.~\ref{fig_stargon}.
We can use this result to compare the interfacial thickness determined by Eq.~\eqref{st_woodmodif} with the simple expression $\xi_0$. Such comparison is shown on Fig.~\ref{fig_xieffcompar3d}. The applicability of the expression is restricted by that of the Guggenheim's law (valid for $0.55 < T/T_c$) and the usage of $\xi_0(T)$ which becomes invalid for low temperatures.
Indeed the deviation of $\xi_0$ from $\xi_{eff}$ becomes noticeable for low temperatures (see Fig.~\ref{fig_xieffcompar3d2}) where the exponential behavior of $n$ on the temperature along the binodal must prevail rather the simple polynomial one. To overcome this difficulty one can use the results of the simulations for the correlational length of the 3D Ising model.

As a summary we have demonstrated that the global isomorphism approach based on simple transformations Eq.~\eqref{projtransfr} can be applied not only for bulk properties of the coexisting phases but also for the description of the surface tension. It opens the way to the application of the results obtained within the lattice models to the description of the real fluids of the LJ type. Since the idea behind Eq.~\eqref{projtransfr} is heavily based on the restoration of the particle-hole symmetry the application to the Tolman length \cite{surftension_tolman_jcp1949} is of special interest. This is due to the fact that this finite size correction to the surface tension vanishes for the symmetrical models \cite{crit_tolmanlengthmpfisher_prb1984}. Therefore it is natural that at least phenomenologically the Tolman length can be directly connected with the density diameter \cite{crit_tolmanlengthanis_prl2007}. Within the global isomorphism approach this means that it is possible to relate the Tolman length with the asymmetry parameter $a$. This will be the subject of the future work.

%%%%%%%%%%%%%%%%%%%%%%%%%%%%%%%%%%%%%%%%%%%%%%%%%%%%%%%%%%%%%%%
%\bibliography{books_my,math,criticality,thermodyn,liqmetals}

\end{document}